# Design of Simple and Efficient Revocation List Distribution in Urban areas for VANET's

Ghassan Samara , Sureswaran Ramadas
National Advanced IPv6 Center, Universiti Sains Malaysia
Penang, Malaysia
ghassan@nav6.org , sures@nav6.org

Wafaa A.H. Al-Salihy
School of Computer Science, Universiti Sains Malaysia
Penang, Malaysia
wafaa@cs.usm.my

*Abstract-* **Vehicular Ad hoc Networks is one of the most challenging research area in the field of Mobile Ad Hoc Networks, in this research we propose a flexible, simple, and scalable design for revocation list distribution in VANET, which will reduce channel overhead and eliminate the use of CRL. Also it will increase the security of the network and helps in identifying the adversary vehicles.**

**We are proposing an idea for using geographical revocation information, and how to distribute it.**

## I. INTRODUCTION

VANET security has gained the most research efforts in the past years; Certificate plays an important role in network communication, any node in the network can't participate in the network without appropriate certificate, certificate distribution is a tough task as a lot of considerations appear like Processing overhead, network overhead, privacy etc.

Dealing with certificate raises other important issue which is certificate revocation list (CRL) that causes network overhead in VANET, a CRL is a list containing the serial numbers of all certificates issued by a given certification authority (CA) that have been revoked and have not yet expired. CRL makes overhead and expensive to use especially in high mobile network.

In this paper we concerned with certificate revocation distribution, how to protect system from adversary vehicles, how to distribute information about adversary vehicles (Revocation List), in sec. 2 we analyze the current research efforts in area of VANET certificates, in sec. 3 we are addressing our proposed network that contains solutions for current system.

## II. ANALYSIS OF RELEVANT RESEARCH AREA:

CRL is the most and common solution for certificate revocation and management, many papers tried to adapt the CRL solutions.

Efforts made by [1] aiming to reduce the CRL by using regional CA and using short lived certificates for traveling vehicles called FC, these certificates must be used in foreign territory and must be tracked and initiated by home CA, this solution needs to be used for large geographical area, like countries, but in this case the CRL will be huge, and if the area is smaller, many CF's will be created, and the tracking will be costly, the result obtained by the author " the distribution of CRL requires tens of minutes", which is too long time for a high and dense network like VANET, the authors in [2] proposed an idea to easy disseminate the CRL, by deploying C2C communication for distributing CRL, this will make faster distribution, but still CRL has a huge size and require time and processing complexity to search in, another work in [3] , made many experiments on the size of CRL and how to distribute the CRL in the VANET network, the result says, when the size of CRL is high, the delay time for receiving it will be high, another idea proposed in [4] says that CRL will store entries for less than one year, this idea used to decrease the size of CRL, but still suffer from huge size, while Authors in [6] suggested a way to increase the search in CRL by using Bloom filter, the problem of bloom filter as it is probabilistic function, and may give wrong information, as the certificate may not be in the list, and the result that the certificate is in the list.

The authors in [7] proposed the use of Bloom filter to store the revoked certificate, and dedicate the CRL just to sign the revocation key for each vehicle, the use for Bloom filter will increase the speed for searching in it, but still the idea is to use the CRL.

The previous work and efforts didn't eliminate the Problems of CRL like Huge size, no central database for it, Channel, communication and processing overhead.

Authors in [7] proposed that each vehicle must be stored with approximately 25000 certificates, if each certificate has 100 bytes; you can imagine the size of CRL when revoking the information for just one vehicle, and how much time required for search in it.

Authors in [9] introduce the use of temporary certificates and credentials for each geographic area so any new vehicle will not find any difficulty for communicating with current network, by creating new certificates and ID's, but this solution requires a dedicated work from CA to create and revoke certificates for each new coming and leaving vehicle.

## III. PROPOSED NETWORK

Each vehicle equipped with Tamper proof device (TPD) which contains a set of private/ public key that must be used for any communication in VANET network, these keys provided by Central Certificate Authority (CCA), each CCA is responsible for number of Local Certificate Authority (LCA) located in cluster [8], Each LCA is responsible for a specific cluster and its Road Side Units (RSU), TPD also contains certificate for ensuring the identity of the vehicle and allowing the vehicle to communicate.

The idea is to use geographical information for certificate, each 4 KM2 will be treated as a cluster, this cluster contains LCA, this LCA has the whole information about it is cluster, and especially the revocation information for all the vehicle that travels in this cluster, this information is collected from RSU located in this cluster, and from another LCA's located in neighbor clusters, see figure 1.

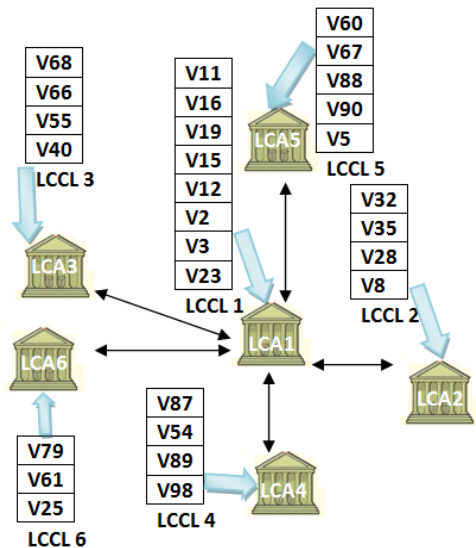

Fig. 1: Local CA Structure.

Each LCA transmit its Local Cluster Certificate List (LCCL) that contains the revoked certificates from its cluster to all nearest RSU's that surround the Local Certificate Authority (LCA), every minute, the list is small as it contains just the revoked certificate from that cluster, so transmission is fast and no transmission overhead is produced.

Each RSU contains two lists, first list is LCCL from LCA of the cluster that RSU belongs to, this list is inserted into every incoming vehicle as a revocation list, second list comes from neighbor LCA and called Neighbor Cluster Certificate List (NCCL), and is used to be searched in by the RSU when new vehicle arrives at the border of the cluster to know if the incoming vehicle is adversary or not.

The size of LCCL is quite small and will not exceed few KB, while the size for CRL holding revocation information for just one vehicle will exceed 2 MB [7]

### A. The protocol:

Each incoming vehicle to the road or to the cluster, will interact with RSU, RSU's will be located in the beginning of each road/ cluster and on important intersections as "cluster guard ", and will be put on traffic light, stop sign or street columns etc., the incoming vehicle must slow its speed, as it crossing an intersection, this will give the RSU the opportunity to interact with the incoming vehicle, the communication is very fast, and happened as follows:

1- RSU takes the Public Key (PK) and certificate of the vehicle.

2- RSU insert its PK, Cluster Signature and updates the vehicle LCCL with the new one for the current cluster.

3- RSU searches for the certificate in NCCL, to make sure that if the vehicle is adversary or not.

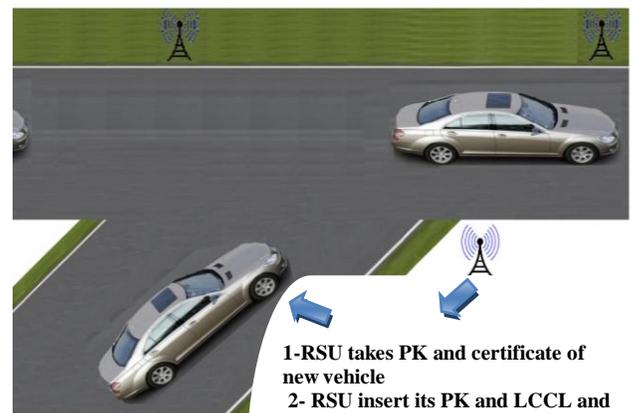

Fig. 2: new vehicle arrives.

**1-RSU takes PK and certificate of new vehicle**
**2- RSU insert its PK and LCCL and Cluster Signature.**
**3- RSU search in NL for vehicle certificate**
**4- if certificate in NCCL inform local and neighbor LCA**

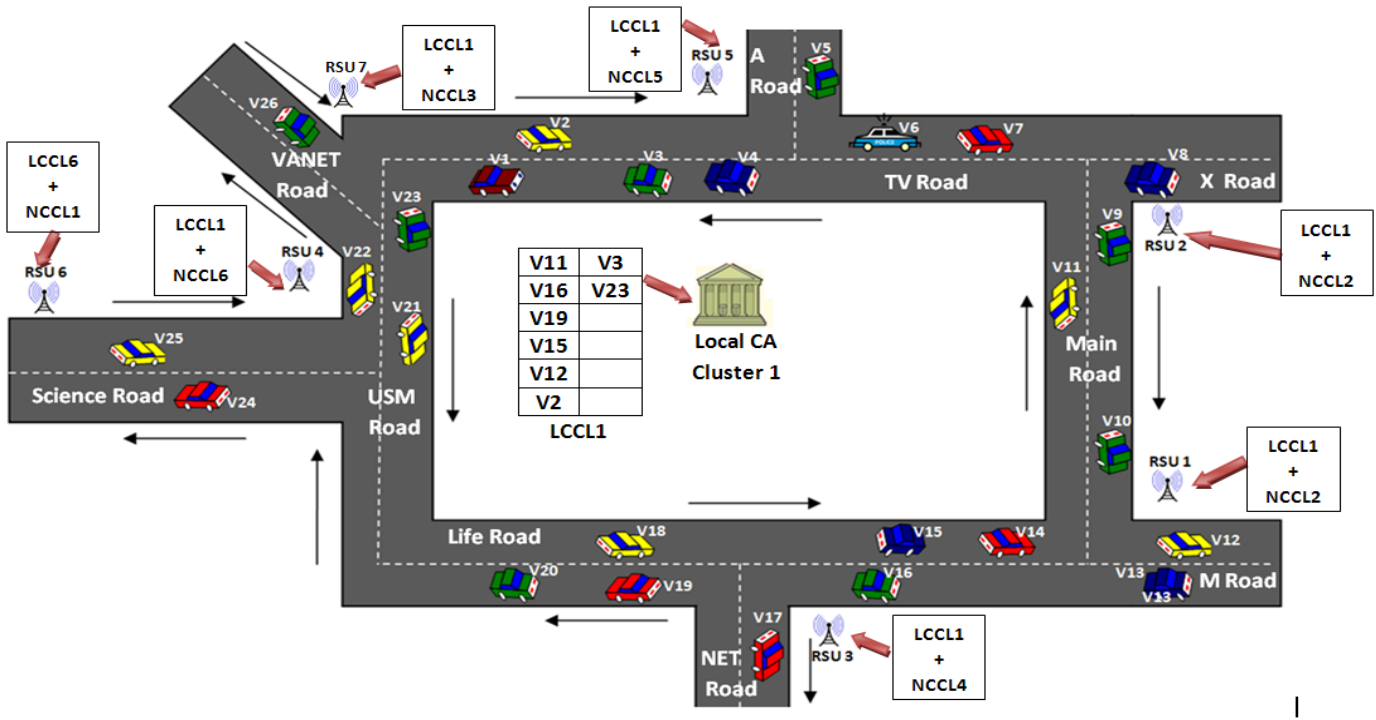

Fig. 3: the Cluster

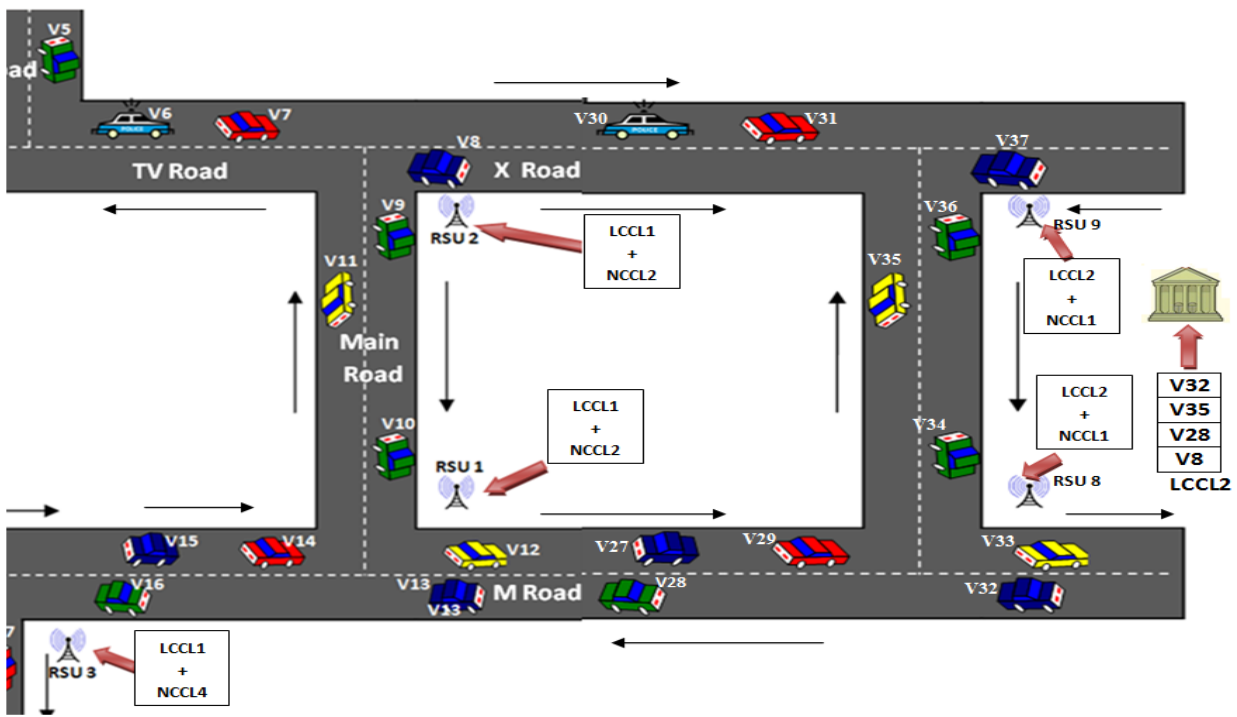

Fig. 4: Neighbor Cluster.

4- If the certificate in NCCL, RSU will sends Add message to current LCA to add this adversary to local LCCL and sends Remove message to neighbor LCA that vehicle come from to remove that vehicle from its LCCL, see figure 2, 3 and 4.

5- Each LCA transmits the LCCL every one minute, but if it receives an Add request from RSU, it will add the certificate to LCCL, LCA may receive more than Add request from more than one RSU in the same cluster, after adding new certificates to LCCL, LCA sends the updated LCCL to whole

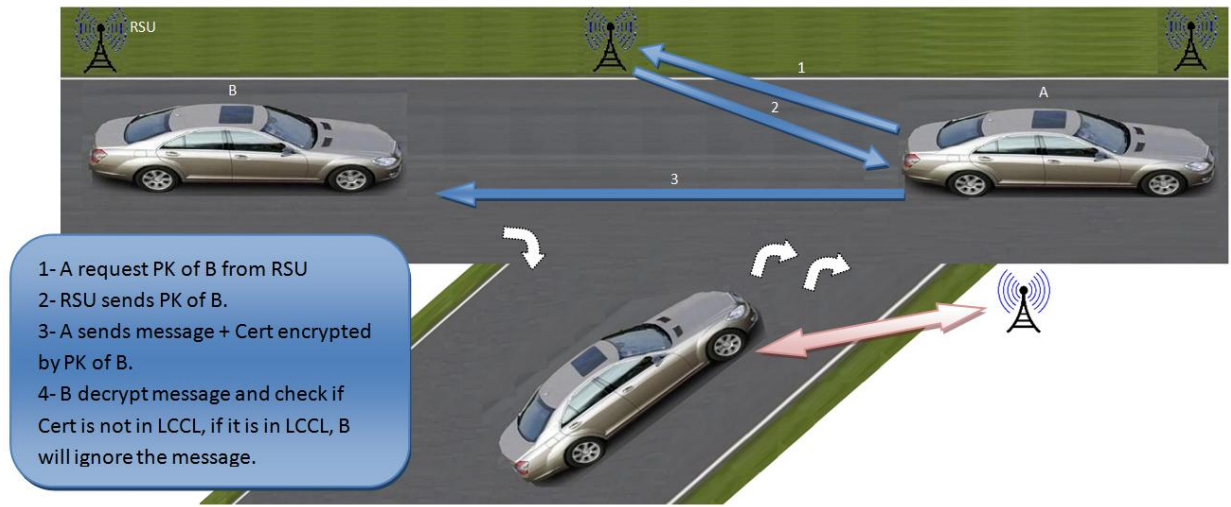

Fig. 5: How the System works.

cluster including local RSU's and vehicle and to neighbor RSU's, and reset the counter for transmission, as the new transmission will start after 1 minute, unless adversary vehicle sited

6- Neighbor LCA removes the certificates of leaving adversary vehicle from its cluster, and change its signature, and sends new signature and new LCCL to whole members of its cluster, changing cluster signature will prevent any adversary vehicle from interact with any vehicle on other clusters.

### A. Car 2 Car Communication:

The communication starts when any vehicle wants to communicate with another vehicle, the sender communicates with nearest RSU asking for the target PK as the message sent must be encrypted, RSU will reply and giving the PK of the target, after the sender get the PK of the target it will start sending the message containing

PKR {M, PKS, SignV, Cert, SignG}

Where M is the message, PK is the public key of the sender, SignV is the sender signature, Cert is the sender certificate, and SignG is the group signature.

All of this information is encrypted with public key of the receiver, after the receiver receives the message it make the following steps:

1. Receiver vehicle decrypt the message with its private key.
2. Detect the correctness of the cluster signature, as the adversary may try to communicate with vehicles of other clusters within its range.
3. Check if the sender is not adversary, by applying its certificate into a function for searching into LCCL, if the certificate is in LCCL, it will not take the message, and it will move the certificate into the top of LCCL, to make the searching in the future faster, as this adversary may try to send again, as it has this vehicle PK. See figure 5.

Searching in LCCL is quit fast and of course faster than searching in CRL, as it contains just the certificate revocation information from local cluster not from the whole world.

### B. Example on figure 3 and 4:

When V25 travels on Science Road it has LCCL6 in its TPD, when it arrives to RSU4 several steps must be made in order to enter new cluster:

1- RSU4 inserts LCCL1 and PK of into vehicle.

2- RSU4 takes certificate and PK of vehicle for further communication.

3- RSU4 search in NCCL6 that it has, and finds that V25 is an adversary.

4- RSU4 sends Add request for LCA1 encrypted by public key of LCA1 and signed by RSU4 signature to add V25 into LCCL1, RSU4 also sends Remove request for LCA6 to remove V25 certificate from LCCL6, Add request has higher priority than Remove request so it must be implemented first.

5- LCA1 receives at the same time Add request from RSU2, where RSU2 founds that V8 in NCCL2, and another Add request from RSU5 for V5 which mentioned in NCCL5.

6- LCA1 updates its list, adding V8, V25, V5 into LCCL1.

7- LCA1 sends the updated LCCL1 for the whole cluster, including local vehicle, local RSU, and neighbor RSU's.

8- LCA1 restart the timer, and will retransmit LCCL1 again after one minute.

LCCL1 after update will contain:

| V8  | V25 | V5  | V11 | V16 |
|-----|-----|-----|-----|-----|
| V19 | V15 | V12 | V2  | V3  |

Table 1: LCCL1 Contents.

### C. Another Considerations

#### i. Grey Area:

Grey areas is the areas that lies between two clusters, and the signals of LCA can't reach it, when a vehicle parking in that area, and tries to communicate with vehicle in other cluster within its range, but it can't as it doesn't have the signature of that cluster, the vehicle can make a request from nearing RSU from that cluster, RSU will check in its NCCL, to make sure that the vehicle is not adversary, if it I not adversary, RSU will give it group signature and LCCL of that cluster.

#### ii. Safety Message:

RSU is responsible for sending the safety message, as it will distribute it for a wide range, and follow up the status of the problem, this will be as follows:

1. RSU takes the safety message from vehicle.
2. RSU sends the message for LCA, to send it for other RSU's.
3. Any incoming vehicle to the cluster, will receive this safety message, and other important message and instruction hints, as the vehicle approaching the cluster, as a (Local Cluster News).

#### iii. Network Size:

The maximum size for the cluster is 4 km$^2$ as DSRC signal range can reach 1 KM, LCA will transmit for 1 KM for each direction, authors in [10] propose the size of the cluster to be 8 KM$^2$, the size has a tradeoff, if it is smaller, it will increase the network overhead, and will require more RSU and LCA to be installed, and if it is bigger it will require multi hop transmission for the information, and the size of LCCL and NCCL will be bigger.

#### iv. Urban Area Characteristics

Normally urban areas are dense in vehicles, buildings and vehicles, roads are smaller than highways, speed limits degrade the mobility, and speed problems that VANET suffer from.

Many Stop signs and Traffic light that will give a chance for installing small RSU's.

These Characteristics will give RSU's a good chance to communicate with vehicles.

### IV. CONCLUSION AND FUTURE WORK

In this paper we introduced new method for distributing revocation information in urban areas for VANETs, as we divided the network into clusters, and we gave LCA the ability to control this cluster, RSU is the cluster guard, which monitor and sense the incoming adversary, and make quick report about it, this method will make the network less overhead, and the communication faster, as we eliminated the use of CRL, in our future work we would like to make simulation for previous protocol and methods.